# Observation of Quantum Griffiths Singularity and Ferromagnetism at Superconducting LaAlO$_3$/SrTiO$_3$(110) Interface

Shengchun Shen,[1,*] Ying Xing,[2,*] Pengjie Wang,[2,*] Haiwen Liu,[2,3] Hai-Long Fu,[2] Yangwei Zhang,[2] Lin He,[1] X. C. Xie,[2,3] Xi Lin,[2,3,†] Jiacai Nie[1,†] and Jian Wang[2,3,†]

[1] *Department of Physics, Beijing Normal University, Beijing 100875, People's Republic of China.*

[2] *International Center for Quantum Materials, School of Physics, Peking University, Beijing 100871, China.*

[3] *Collaborative Innovation Center of Quantum Matter, Beijing, China.*

\* These authors contributed equally to this work

† **E-mail:** jianwangphysics@pku.edu.cn(J.W.), jcnie@bnu.edu.cn(J.N.) and xilin@pku.edu.cn(X.L.).

Diverse phenomena emerge at the interface between band insulators LaAlO$_3$ and SrTiO$_3$, such as superconductivity and ferromagnetism, showing an opportunity for potential applications as well as bringing fundamental research interests. Particularly, the two-dimensional electron gas formed at LaAlO$_3$/SrTiO$_3$ interface offers an appealing platform for quantum phase transition from a superconductor to a weakly localized metal. Here we report the superconductor-metal transition in superconducting two-dimensional electron gas formed at LaAlO3/SrTiO3(110) interface driven by a perpendicular magnetic field. Interestingly, when approaching the quantum critical point, the dynamic critical exponent is not a constant but a diverging value, which is a direct evidence of quantum Griffiths singularity raised from quenched disorder at ultralow temperatures. Furthermore, the hysteretic property of magnetoresistance was firstly observed at LaAlO$_3$/SrTiO$_3$(110) interfaces, which suggests potential coexistence of superconductivity and ferromagnetism.

Two dimensional electron gas (2DEG) formed at the interface between two band insulators LaAlO$_3$ (LAO) and SrTiO$_3$(STO) [1] exhibits many fascinating properties, such as superconductivity [2], magnetism [3], and their coexistence [4-6]. Historically, 2DEG has been a perfect system to study quantum Hall effect, fractional quantum Hall effect, charge density wave and the transition between them by varying charge density or magnetic field [7-10]. In 2DEG of LAO/STO(001), quantum phase transition (QPT) is also an important topic and transition from superconducting state to weakly insulating state was studied [11,12]. Most previous studies are focused on LAO/STO(001) due to feasible fabrication and the polar discontinuity at the interface. Recently, controlled growth on pseudo-cubic orientation LAO/STO(110) non-polarized interfaces has been successful and made the related investigations possible [13-15].

Commonly, the critical behavior depends only on the universality class of the transition, and not on microscopic details. However, quenched disorder can have profound effects on phase transitions and critical points, especially in QPTs [16-18]. If the average disorder strength diverges under coarse graining, the critical behavior shows infinite-randomness quantum critical points (QCPs) where the conventional power-law scaling is replaced by the activated scaling, namely the quantum Griffiths singularity [19-22]. Quantum Griffiths singularity has been widely studied in theoretical works, but with limited experimental evidences in three dimensional ferromagnetic metals [17,23]. Surprisingly, a recent unexpected observation exhibited the quantum Griffith singularity of superconductor-metal transition (SMT) in three monolayers superconducting Ga films [24,25]. Whether or not the quantum Griffiths singularity is a universal scenario for QPTs in superconducting systems still manifests as a very interesting question. The verification of the quantum Griffiths singularity can help to identify the role of disorder in superconducting systems and hopefully may provide clues to a generalized theoretical framework for disordered QPTs.

Here we report the observation of quantum Griffiths singularity at superconducting LAO/STO(110) 2DEG interface. QPT from a superconductor to a weakly metal is driven by perpendicular magnetic field. The conventional power-law scaling is replaced by activated scaling, and the dynamical exponent $z$ diverges upon approaching the transition. An unconventional quantum critical behavior associated to quantum Griffiths singularity is deduced and infinite randomness quantum critical points are further concluded. Additional data with similar behaviors for backgate voltage $V_G$ = 20 V and 60 V at presented for a comprehensive understanding. Besides, the hysteretic magnetoresistance is observed, which indicates the existence of ferromagnetism at LAO/STO(110) interface.

5 unit cells of LAO films were grown by Pulsed Laser Deposition (PLD) (KrF, $\lambda$ = 248 nm, laser fluence 1.5 J/cm$^2$, 1 Hz, at 750℃), on top of treated (110) STO substrates (see Supplemental Material [26]). Atomically flat (110)-SrTiO$_3$ surfaces were obtained after annealing at 1050℃ for 2 h under oxygen atmosphere. Before deposition, the substrate was heated from room temperature to 750℃ in 0.1 mbar of O$_2$, and then the LAO layer were deposited in $10^{-5}$ mbar of $O_2$. After deposition, the sample was cooled in oxygen rich atmosphere to avoid the formation of oxygen vacancy. More details were described in Ref. 14.

Superconductivity and the superconductor-metal transition are found in our LAO/STO(110) samples. As shown in Fig. 1(a), the temperature dependence of resistance at zero magnetic field reveals the interface is superconducting with superconducting transition temperature $T_c^{zero} \approx 0.123\ K$ and $T_c^{onset} \approx 0.711\ K$, respectively. $T_c^{zero}$ is identified as the temperature at which the $R$ drops beyond the measurement limit, while $T_c^{onset}$ is identified as the temperature where the $R(T)$ firstly deviated from its linear dependence at high temperature [see the inset of Fig. 1(a)]. With increasing magnetic field, the superconductivity is suppressed, the system gradually becomes weakly insulating. The isomagnetics $R(T)$ shows that resistance changes very slightly at the ultralow temperatures with a critical field (around 0.417 T) that separates two regimes, as shown in Fig. 1(b). As the signature of $B$-driven SMT, the magnetoresistance $R(B)$ measured at different temperatures crosses each other around 0.385 T, as shown in Fig. 1(c). Previous studies have shown a SIT occurs in LAO/STO(001) system induced by magnetic field or charge density [11,12].

The crossing of the magnetoresistance isotherms happens in a well distinguished region rather than a single point in the magnetic field, as shown in Fig. 1(c). Similar crossing region was also observed in 12-unit-cell-LAO/STO(110) sample [see Fig. S5 in Supplemental Material [26]]. The information of crossing allows a systematically investigation of the critical behavior, which was done through careful measurements of the $R$ versus $B$ in temperature from 0.020 K to 0.650 K. As shown in Fig. 2, a series of crossing points are observed. Crossing points of $R(B)$ isotherms at every two adjacent temperatures as a function of $T$ is shown in the inset of Fig. 2 (blue dots), which form a roughly linear line. The $R$ plateaus extracted from $R(T)$ curves are shown as green squares, at which the $dR/dT$ changes sign for a given magnetic field. The crossing points are well consistent with $R$ plateaus. Besides, we note that the crossing region of magnetoresistance curves extends over a relatively wide range of magnetic field and temperature. It is reasonable since in our LAO/STO(110) samples superconductivity emerges at relatively high temperature [$T_c^{onset} \approx 0.711$ K, inset of Fig. 1(a)].

In order to gain more information about the phase transition, the finite-size scaling (FSS) analyzing procedure of the critical regime is required. The resistance takes from the scaling form [24,27-29],

$$R(B,T) = R_c f[(B - B_C)/T^{1/zv}], \qquad (1)$$

where $R_C$, $B_C$ are the critical sheet resistance and the critical magnetic field respectively at which the transition occurs, $f[\,]$ is an arbitrary function of $B$ and $T$ with $f[0] = 1$, $z$ is the dynamical critical exponent, $v$ is the correlation length exponent. The scaling form is rewritten as $R(B,t)/R_C = f[(B - B_C)t]$, where $t = (T/T_0)^{-1/zv}$ can be obtained by performing a numerical minimization between the curve $R(B,t)$ at a particular temperature $T$ and the lowest temperature $T_0$ curve $R(B, t = 1)$. The small crossing region, formed by adjacent four $R(B)$ curves, is regarded approximately as one "critical" point. Fig. S2(a) (see Supplemental Material [26]) shows one representative group of isothermal curves with a "critical" point ($B_c$ = 0.427 T, $R_c$ = 877.50 Ω). The results of FSS analysis show the data collapse onto a bi-value curve, and the power law dependence of $t$ with temperature gives $zv$ = 3.37±0.50 [Fig. S2(b) in Supplemental Material [26]]. Nine representative groups with different temperature regions were selected. And a series of $zv$ values can be obtained. It has been reported [30] that FSS analysis can be applied in a restricted range of finite temperature for 2D superconductors under magnetic field, if considering the existence of inhomogeneities in low temperature phase.

Figure 3 shows the magnetic field dependence of $zv$ values. In high temperature regime, $zv$ increases slowly with magnetic field, while $zv$ grows rapidly in low temperature regime. We fit the experimental $zv$ values ($zv \geq 1$) as a function of $B$ by the activated scaling law $zv \approx C|B - B_c^*|^{-v\psi}$, where $C$ is constant, $v \approx 1.2$ and $\psi \approx 0.5$ are the 2D infinite randomness critical exponents [31,32]. The experimental results are in reasonable agreement with theoretical expectation (the blue line in Fig. 3, where $B_C^* = 0.428\ T$), indicating infinite randomness quantum critical point. As decreasing temperature, the effect of quenched disorder is enhanced, the $zv$ value diverges when the critical point $B_c^*$ is approached.

Moreover, the properties of 2DEG at the LAO/STO interface can be tuned by a backgate voltage. Figure 4(a) shows a $R(T)$ curve at zero magnetic field for $V_G = 20$ V with the superconducting transition temperature $T_C^{zero} \approx 0.109\,K$ and $T_C^{onset} \approx 0.696\,K$, which are a little suppressed comparing with those at zero back gate voltage ($T_c^{zero} \approx 0.123\,K$ and $T_c^{onset} \approx 0.711\,K$). The $R(B)$ isotherms show a crossing region around 0.365 T [Fig. 4(b)]. Further careful measurements of magnetoresistance at low temperature range from 0.020 K to 0.300 K are shown in Fig. 4(c). A series of crossing points formed by every two adjacent magnetoresistance isotherms are observed. The extracted $B_c$-$T$ plot is shown in the inset of Fig. 4(c), which is similar to the above state for $V_G = 0$ V [the inset of Fig. 2]. The same FSS analysis described above is applied, then a series of $zv$ values are obtained. As shown in Fig. 4(d), the $zv$ value diverges when temperature approaching zero, theoretical fitting gives $B_c^* = 0.416$ T. The SMT for both $V_G = 20$ V and 60 V [Fig. S4 in Supplemental Material [26]] displays behaviors that are also consistent with quantum Griffith singularity. Quenched disorder is independent of time, it can take the form of oxygen vacancies or impurity atoms, or extended defects and so on. Although the properties (such as carrier density or $T_c$) of 2DEG at LAO/STO interface can be tuned by applying a gate voltage, the impact of quenched disorder on phase transitions still exists, thus the quantum Griffiths singularity for different gate voltages can be observed. Besides, compared to state for $V_G = 0$ V, as increasing $V_G$, $T_c$ decreases and the $zv$ value for the same temperature region becomes larger, which suggests the influence of disorder on phase transition might be tuned by gate voltages.

The quenched disorder is normal in realistic low dimensional systems and plays an important role in the destruction of clean critical point [16,17]. From a statistical point of view, the influence of randomness on a clean critical point is determined by the trends of average disorder strength under coarse graining, known as the Harris criterion [33]. A recent theoretical work [34] has proposed that under certain conditions if the clean critical point violates the Harris criterion, the magnitude of inhomogeneities increases without limit under coarse graining and the dynamical exponent z diverges when approaching the disordered critical point. This prediction connects the violation of Harris criterion with the existence of quantum Griffiths singularity, and can apply to a variety of systems including the quantum random transverse field Ising model [21,22,35]. In recent, some experimental signatures related to quantum Griffiths singularity been reported [36-41]. On the microscopic level, this quenched disorder introduces large rare regions, which can locally ordered in one phase and further influence the scaling behavior. The above mentioned theoretical results can also be applied to SMT systems with clean critical exponent $v = 1/2$, in which the Harris criterion is certainly violated under quenched disorder [34,42].

For the SMT at the LAO/STO(110) interface, we find the critical exponent $zv$ diverges at the ultralow temperature limits and we attribute such an unconventional quantum critical behavior to the effect of quenched disorder [16,17,21,22,43]. In the superconducting 2DEG at LAO/STO(110) interface, there is a transition from a clean ($zv \leq 1$) to a dirty limit [Fig. 3]. The scaling exponent $zv$ diverges rapidly upon approaching QCP, which is consistent with the activated scaling behavior. When approaching the quantum critical point, the effect of quenched disorder overtakes the thermal fluctuations, and results in large local superconducting islands in which the dynamics is frozen. At low temperatures, these superconducting islands couple each other via long-range Josephson coupling [44], thus global superconductivity emerges. While at high temperature, the thermal fluctuations smear the inhomogeneity induced by quenched disorder. In all, the accordance of experimental observation and theoretical expectation suggests the SMT at LAO/STO interface is of the infinite randomness type.

Recently, the direct evidence of quantum Griffiths singularity was observed in a thin film superconducting system [24]. In Ga film, an anomalous upturn of upper "critical field" when approaching zero temperature was observed. While for our LAO/STO interface, we did not observe the pronounced "tail" at ultralow temperatures [inset of Fig. 2]. One possible reason for the absence of this extended phase boundary is due to the ultralow superconducting transition temperature of this interface superconductor, which is one order of magnitude lower than that of Ga film (~3.62 K). In Ga film, the temperature at which the critical exponent diverges is ultralow (~0.075 K) [24]. If simply comparing the results in LAO/STO interface with that of Ga film, the pronounced "tail" should be expected to emerge at lower temperature (~0.025 K). Actually, a crossover-like behavior is observed at 0.030 K for our LAO/STO interface. However, the lowest electron temperature we had is 0.020 K. Hence, it is hard to observe the obvious extended phase boundary at the LAO/STO(110) interface.

Quantum criticality in magnetic field driven QPT has been studied previously at superconducting oxide interface, such as LaTiO$_3$/SrTiO$_3$(001) and LAO/STO(001) interfaces [12,45]. In LaTiO$_3$/SrTiO$_3$(001) interface, two different $zv$ values were obtained, and the observed critical behavior is single or double which depends on its conductance [45]. While in LAO/STO(001) interface, single $zv$ value was obtained which is independent of its conductance [12]. Here, however, we found a series of $zv$ values, independent of its conductance. In the low temperature range, $zv$ value is larger. With increasing temperature, $zv$ value decreases. Moreover, we note that the resistance range of crossing region only extends over a few tens of ohm. Thus, for lower conductance, the crossing region is hard to distinguish as Ref. 43. Ultralow temperature fine resolution measurements are necessary to observe the Griffiths singularity in such interface superconducting systems.

Moreover, it is worth mentioning that the ferromagnetism was observed at LAO/STO(110) interface. As shown in Fig. 5, when sweeping the field in both directions, hysteretic magnetoresistance is observed. The main peaks in the magnetoresistance appear at $B \approx \pm 0.06$ T and are weakened with increasing temperature. Besides, less prominent peaks appear around main peaks. The hysteresis is reminiscent of the presence of ferromagnetism order and the additional peaks can be attributed to more complex magnetization dynamics [46]. Moreover, the amplitude of peaks decreases when decreasing the field sweep rate [Fig. S7 in Supplemental Material [26]]. Similar hysteretic magnetoresistance is observed at the superconducting LAO/STO(001) interfaces as the transport evidence for coexistence of superconductivity and ferromagnetism [5,46]. The measurements of scanning superconducting quantum interference device [4] and magnetic torque magnetization [6] also evidenced the ferromagnetism at superconducting LAO/STO(001) interfaces. However, to our knowledge, at LAO/STO(110) interface, the coexistence of superconductivity and ferromagnetism has not been reported yet. Superconductivity and ferromagnetism are usually believed to be antagonistic phenomena. Two scenarios have been proposed to explain coexistence of superconductivity and ferromagnetism at LAO/STO(001) interface. One is the unconventional pairing mechanism, such as finite momentum pairing [47], through which a magnetic ordering and superconducting 2DEG is formed by the same electron system [48]. The other scenario is the spatial phase separation, in which magnetism and superconductivity is generated by different electron layers [4,49]. Our current results cannot distinguish which mechanism induces the coexistence of superconductivity and ferromagnetism at LAO/STO(110) interface. Since the (110) interface shows different orbital reconstruction compared with (001) system [50], our observation offers a new platform to study superconductivity and ferromagnetism at the oxides interface. It would be a very interesting topic whether or not the existence of ferromagnetism at superconducting (110) interface does affect the observed quantum Griffiths singularity, which deserves further theoretical and experimental investigations.

In conclusion, we have shown a SMT in the superconducting 2DEG at (110)-orientation LAO/STO interface. The diverging dynamic critical exponent is consistent with quantum Griffiths singularity. Diverging dynamic critical in 2DEG provides new evidence of quantum Griffiths singularity in addition to that in Ga thin film, hinting that different superconducting systems can be possibly treated within a uniform theoretical framework. Furthermore, our detection of hysteretic behavior indicates the ferromagnetism at superconducting LAO/STO(110) interfaces.

## Acknowledgements

We thank Fa Wang and Limei Xu for helpful discussions. We acknowledge financial support by National Basic Research Program of China (Grant Nos. 2013CB934600, 2012CB921300, 2013CB921701, 2013CBA01603, 2015CB921101, and 2014CB920903), the National Natural Science Foundation of China (Grants No. 11474022, No. 51172029, No. 11374035, No. 11274020, No. 11322435 and No. 11422430), and the Research Fund for the Doctoral Program of Higher Education (RFDP) of China. S.S., Y.X. and P.W. contributed equally to this work.

## References

[1]  A. Ohtomo and H. Y. Hwang, Nature **427**, 423 (2004).
[2]  N. Reyren, S. Thiel, A. D. Caviglia, L. F. Kourkoutis, G. Hammerl, C. Richter, C. W. Schneider, T. Kopp, A. S. Ruetschi, D. Jaccard, M. Gabay, D. A. Muller, J. M. Triscone, and J. Mannhart, Science **317**, 1196 (2007).
[3]  A. Brinkman, M. Huijben, M. Van Zalk, J. Huijben, U. Zeitler, J. C. Maan, W. G. van der Wiel, G. Rijnders, D. H. Blank, and H. Hilgenkamp, Nat. Mater. **6**, 493 (2007).
[4]  J. A. Bert, B. Kalisky, C. Bell, M. Kim, Y. Hikita, H. Y. Hwang, and K. A. Moler, Nat. Phys. **7**, 767 (2011).
[5]  D. Dikin, M. Mehta, C. Bark, C. Folkman, C. Eom, and V. Chandrasekhar, Phys. Rev. Lett. **107**, 056802 (2011).
[6]  L. Li, C. Richter, J. Mannhart, and R. C. Ashoori, Nat. Phys. **7**, 762 (2011).
[7]  K. Klitzing, G. Dorda, and M. Pepper, Phys. Rev. Lett. **45**, 494 (1980).
[8]  R. B. Laughlin, Phys. Rev. Lett. **50**, 1395 (1983).
[9]  D. C. Tsui, H. L. Stormer, and A. C. Gossard, Phys. Rev. Lett. **48**, 1559 (1982).
[10] H. L. Stormer, Rev. Mod. Phys. **71**, 875 (1999).
[11] A. D. Caviglia, S. Gariglio, N. Reyren, D. Jaccard, T. Schneider, M. Gabay, S. Thiel, G. Hammerl, J. Mannhart, and J. M. Triscone, Nature **456**, 624 (2008).
[12] M. Mehta, D. Dikin, C. Bark, S. Ryu, C. Folkman, C. Eom, and V. Chandrasekhar, Phys. Rev. B **90**, 100506 (2014).
[13] G. Herranz, F. Sanchez, N. Dix, M. Scigaj, and J. Fontcuberta, Sci. Rep. **2**, 758 (2012).
[14] Y. L. Han, S. C. Shen, J. You, H. O. Li, Z. Z. Luo, C. J. Li, G. L. Qu, C. M. Xiong, R. F. Dou, L. He, D. Naugle, G. P. Guo, and J. C. Nie, Appl. Phys. Lett. **105**, 192603 (2014).
[15] G. Herranz, G. Singh, N. Bergeal, A. Jouan, J. Lesueur, J. Gazquez, M. Varela, M. Scigaj, N. Dix, F. Sanchez, and J. Fontcuberta, Nat. Commun. **6**, 6028 (2015).
[16] T. Vojta, J. Phys. a-Math. Gen. **39**, R143 (2006).
[17] T. Vojta, J. Low Temp. Phys. **161**, 299 (2010).
[18] A. Del Maestro, B. Rosenow, J. A. Hoyos, and T. Vojta, Phys. Rev. Lett. **105**, 145702 (2010).
[19] R. B. Griffiths, Phys. Rev. Lett. **23**, 17 (1969).
[20] B. M. McCoy, Phys. Rev. Lett. **23**, 383 (1969).
[21] D. S. Fisher, Phys. Rev. Lett. **69**, 534 (1992).
[22] D. S. Fisher, Phys. Rev. B **51**, 6411 (1995).
[23] M. Brando, D. Belitz, F. Grosche, and T. Kirkpatrick, arXiv 1502.02898, 1502.02898 (2015).
[24] Y. Xing, H.-M. Zhang, H.-L. Fu, H. Liu, Y. Sun, J.-P. Peng, F. Wang, X. Lin, X.-C. Ma, Q.-K. Xue, J. Wang, and X. C. Xie, Science **350**, 542 (2015).
[25] N. Markovic, Science **350**, 509 (2015).
[26] See Supplemetal Material for details of thin-film preparation and measurements. Additional results are also provided.
[27] M. P. Fisher, Phys. Rev. Lett. **65**, 923 (1990).
[28] A. Goldman, Int. J. Mod. Phys. B **24**, 4081 (2010).
[29] S. L. Sondhi, S. M. Girvin, J. P. Carini, and D. Shahar, Rev. Mod. Phys. **69**, 315 (1997).
[30] N. Mason and A. Kapitulnik, Phys. Rev. B **64**, 060504 (2001).
[31] T. Vojta, A. Farquhar, and J. Mast, Phys. Rev. E **79**, 011111 (2009).
[32] I. A. Kovács and F. Iglói, Phys. Rev. B **82**, 054437 (2010).
[33] A. B. Harris, Journal of Physics C: Solid State Physics **7**, 1671 (1974).
[34] T. Vojta and J. A. Hoyos, Phys. Rev. Lett. **112**, 075702 (2014).
[35] O. Motrunich, S. C. Mau, D. A. Huse, and D. S. Fisher, Phys. Rev. B **61**, 1160 (2000).
[36] C. Seaman, M. Maple, B. Lee, S. Ghamaty, M. Torikachvili, J.-S. Kang, L. Liu, J. Allen, and D. Cox, Phys. Rev. Lett. **67**, 2882 (1991).

[37]  M. De Andrade, R. Chau, R. Dickey, N. Dilley, E. Freeman, D. Gajewski, M. Maple, R. Movshovich, A. C. Neto, and G. Castilla, Phys. Rev. Lett. **81**, 5620 (1998).
[38]  S. Guo, D. Young, R. Macaluso, D. Browne, N. Henderson, J. Chan, L. Henry, and J. DiTusa, Phys. Rev. Lett. **100**, 017209 (2008).
[39]  J. Sereni, T. Westerkamp, R. Küchler, N. Caroca-Canales, P. Gegenwart, and C. Geibel, Phys. Rev. B **75**, 024432 (2007).
[40]  T. Westerkamp, M. Deppe, R. Kuchler, M. Brando, C. Geibel, P. Gegenwart, A. P. Pikul, and F. Steglich, Phys. Rev. Lett. **102**, 206404 (2009).
[41]  S. Ubaid-Kassis, T. Vojta, and A. Schroeder, Phys. Rev. Lett. **104**, 066402 (2010).
[42]  S. Sachdev, P. Werner, and M. Troyer, Phys. Rev. Lett. **92**, 237003 (2004).
[43]  D. S. Fisher, M. P. Fisher, and D. A. Huse, Phys. Rev. B **43**, 130 (1991).
[44]  V. M. Galitski and A. I. Larkin, Phys. Rev. Lett. **87**, 087001 (2001).
[45]  J. Biscaras, N. Bergeal, S. Hurand, C. Feuillet-Palma, A. Rastogi, R. C. Budhani, M. Grilli, S. Caprara, and J. Lesueur, Nat. Mater. **12**, 542 (2013).
[46]  M. M. Mehta, D. A. Dikin, C. W. Bark, S. Ryu, C. M. Folkman, C. B. Eom, and V. Chandrasekhar, Nat. Commun. **3**, 955 (2012).
[47]  K. Michaeli, A. C. Potter, and P. A. Lee, Phys. Rev. Lett. **108**, 117003 (2012).
[48]  M. S. Scheurer and J. Schmalian, Nat. Commun. **6**, 6005 (2015).
[49]  S. Banerjee, O. Erten, and M. Randeria, Nat. Phys. **9**, 625 (2013).
[50]  D. Pesquera, M. Scigaj, P. Gargiani, A. Barla, J. Herrero-Martín, E. Pellegrin, S. Valvidares, J. Gázquez, M. Varela, and N. Dix, Phys. Rev. Lett. **113**, 156802 (2014).

**Figures**

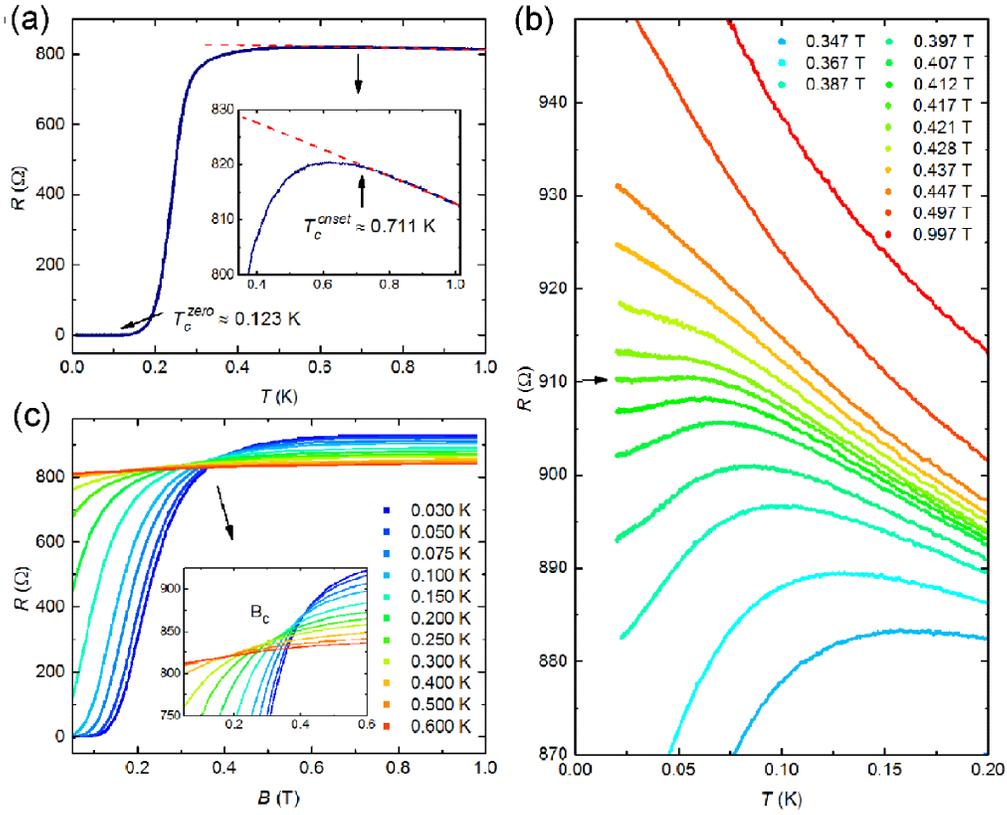

**FIG. 1.** The magnetic field driven superconductor-metal transition. (a) The temperature dependence of resistance at zero magnetic field. $T_c^{zero}$ and $T_c^{onset}$ marked by black arrow is 0.123 K and 0.711 K, respectively. The inset shows the determination of $T_c^{onset}$. (b) Isomagnetic $R(T)$ curves measured at different $B$. (c) Isotherms $R(B)$ curves measured at different $T$, the inset shows the zoom-in view of crossing region.

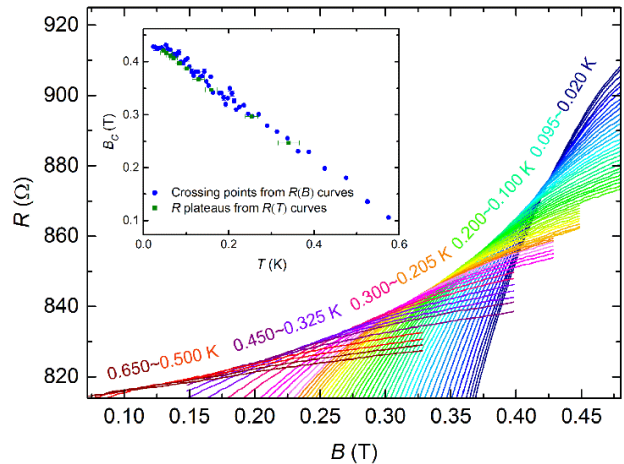

**FIG. 2.** The isotherms of magnetoresistance $R(B)$ from 0.020 to 0.650 K. The sweep directions and rates are the same for all the curves. The inset shows the $T$ dependence of corresponding $B_c$ for crossing points (blue dots) of every two adjacent $R(B)$ curves, and the resistance plateaus (green squares) extracted from $R(T)$ curves.

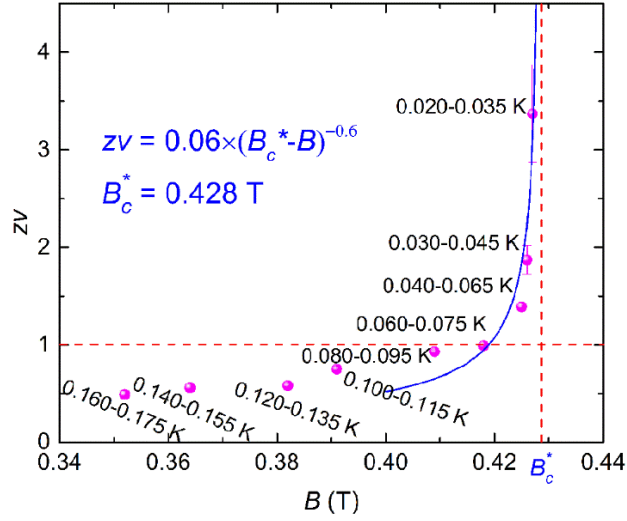

**FIG. 3.** The $B$ dependence critical exponent $zv$: activated scaling behavior. Magenta dots are $zv$ values extracted from FSS analysis in different temperature regions. Blue line is the fitting by $zv = C(B_c^* - B)^{-0.6}$. Two red dash lines represent the constant value with $B_c^*$=0.428 T and $zv$ =1, respectively.

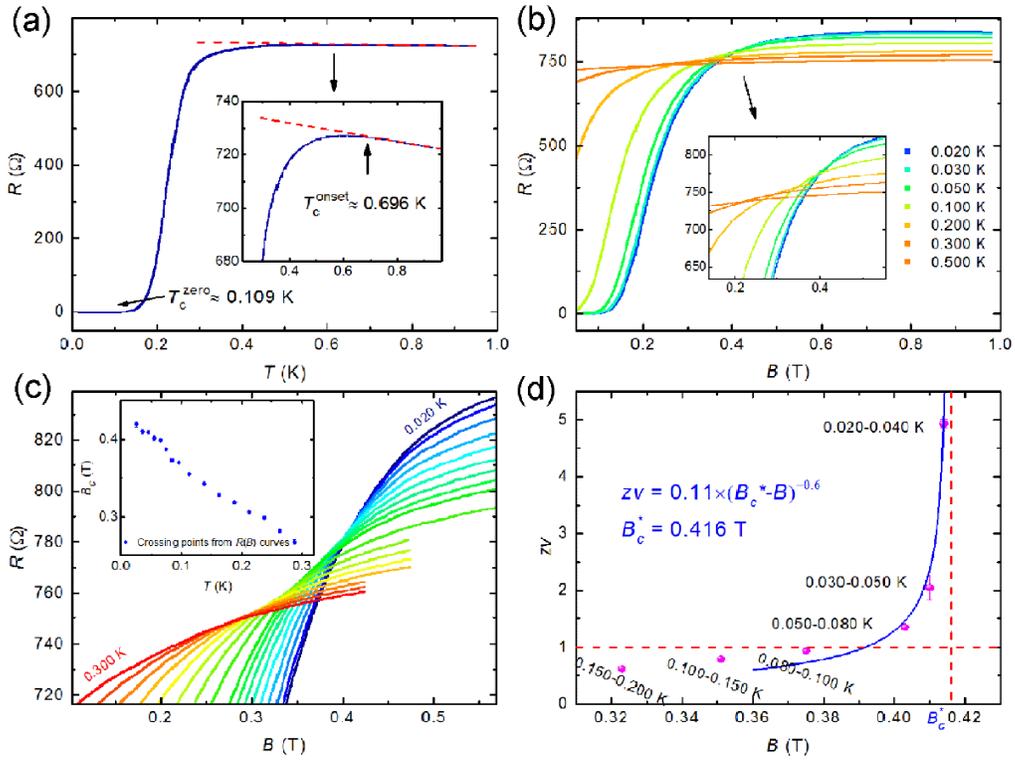

**FIG. 4.** The quantum Griffiths singularity for $V_G$ = 20 V. (a) The $R(T)$ at zero magnetic field with $T_c^{zero}$=0.109 K, and the inset shows the definition of $T_c^{onset}$, with a value of 0.696 K. (b) The isotherms $R(B)$ measured at different $T$. Zoom-in view of cross region is shown in the inset. (c) The isotherms $R(B)$ measured at different $T$ ranged from 0.020 K to 0.300 K. The inset provides the crossing points $B_c$ as a function of $T$, which are determined from the cross point of every adjacent two $R(B)$ curves. (d) The $B$ dependence of $zv$ values reveals the activated scaling behavior.

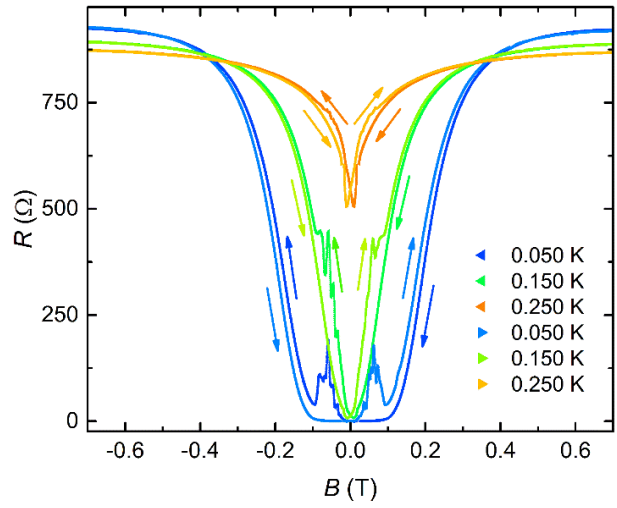

**FIG. 5.** The hysteretic behavior of *R*(*B*) at different temperatures are plotted for $V_G = 0$ V. The peaks emerged in the magnetoresistance are weakened with increasing temperature, while the corresponding magnetic field value keeps unchanged. The arrows indicate the sweep directions of the magnetic field.